\documentclass[a4paper,10pt,twoside]{cpc-hepnp}

\usepackage{multicol}
\usepackage{graphicx}
\usepackage{epstopdf}
\usepackage{subfigure}
\usepackage{float}
\usepackage{array}
\usepackage{booktabs}
\usepackage{amssymb,bm,mathrsfs,bbm,amscd}
\usepackage[tbtags]{amsmath}
\usepackage{lastpage}
\usepackage{CJK}

\begin{document}
\begin{CJK*}{GBK}{song}

\fancyhead[c]{\small Chinese Physics C~~~Vol. xx, No. x (201x) xxxxxx}
\fancyfoot[C]{\small 010201-\thepage}

\title{Study of cluster reconstruction and track fitting algorithms for CGEM-IT at  BES\uppercase\expandafter{\romannumeral3} \thanks{Supported by National Key Basic Research Program of China (2015CB856700), National Natural Science Foundation of China (11205184, 11205182) and Joint Funds of National Natural Science Foundation of China (U1232201)}}

\author {GUO Yue$^{1,2;1)}$\email{guoy@ihep.ac.cn}
\quad WANG Liang-Liang$^{1;2)}$\email{llwang@ihep.ac.cn}
\quad JU Xu-Dong$^{1}$
\quad WU Ling-Hui$^{1}$  \\
\quad XIU Qing-Lei$^{1}$
\quad WANG Hai-Xia$^{3}$
\quad DONG Ming-Yi$^{1}$
\quad HU Jing-Ran$^{3}$   \\
\quad LI Wei-Dong$^{1}$   
\quad LI Wei-Guo$^{1}$
\quad LIU Huai-Min$^{1}$
\quad OU-YANG Qun$^{1}$    \\
\quad SHEN Xiao-Yan$^{1}$  
\quad YUAN Ye$^{1}$
\quad ZHANG Yao$^{1}$  
}
\maketitle

\address{%
$^1$ Institute of High Energy Physics, Chinese Academy of Sciences, Beijing 100049, China   \\
$^2$ University of Chinese Academy of Sciences, Beijing 100049, China \\
$^3$ Central China Normal University, Wuhan 430079, China
}

\begin{abstract}
Considering the aging effects of existing Inner Drift Chamber (IDC) of BES\uppercase\expandafter{\romannumeral3}, a GEM based inner tracker is proposed to be designed and constructed as an upgrade candidate for IDC. This paper introduces a full simulation package of CGEM-IT with a simplified digitization model, describes the development of the softwares for cluster reconstruction and track fitting algorithm based on Kalman filter method for CGEM-IT. Preliminary results from the reconstruction algorithms are obtained using a Monte Carlo sample of single muon events in CGEM-IT.
\end{abstract}



\footnotetext[0]{\hspace*{-3mm}\raisebox{0.3ex}{$\scriptstyle\copyright$}2015
Chinese Physical Society and the Institute of High Energy Physics
of the Chinese Academy of Sciences and the Institute
of Modern Physics of the Chinese Academy of Sciences and IOP Publishing Ltd}%

\begin{multicols}{2}

\section{Introduction}

The Beijing Spectrometer \uppercase\expandafter{\romannumeral3} (BES\uppercase\expandafter{\romannumeral3}) \cite{1} is a detector of high precision that operates at the Beijing Electron-Positron Collider \uppercase\expandafter{\romannumeral2} (BEPC\uppercase\expandafter{\romannumeral2}) \cite{2}, a high luminosity, multi-bunch e$^{+}$ e$^{-}$ collider running at the $\tau$-charm energy region. The Main Drift Chamber (MDC) of BES\uppercase\expandafter{\romannumeral3} is a key subdetector for charged track reconstruction with good spatial and momentum resolutions. MDC comprises two parts, Inner Drift Chamber (IDC) and Outer Drift Chamber (ODC). However, the existing IDC is suffering from aging problems after five years of operation, which results in an obvious decrease in its gas gain and spatial resolution \cite{3}.

\begin{figure}[H]
	\centering	
	\includegraphics[width=0.5\textwidth]{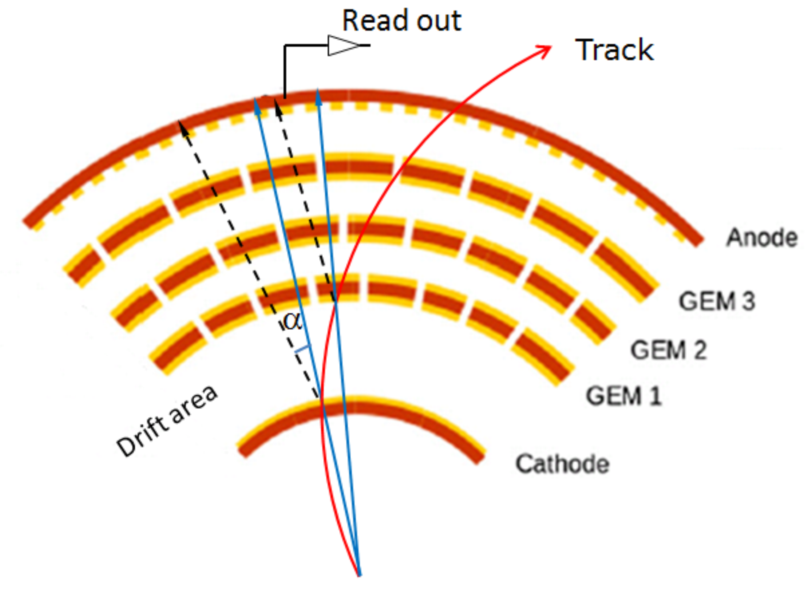}	
	\caption{The structure of one CGEM detector that consists of one cathode, three GEM foils and one anode. When a track (red solid curve) traverses one CGEM, ionization occurs in the drift area (the gap between the cathode and the first GEM foil). The electrons are multiplied in the holes of GEM foils and drift towards the readout anode. The position in anode plane is obtained by digitization with (black dashed lines) and without (blue solid lines) taking into account the Lorentz angle $\alpha$ and electron diffusion.} \label{1.1}
\end{figure}

Gas Electron Multiplier (GEM) detector\cite{4} is characterized by a significant radiation tolerance, high rate capability, an outstanding spatial resolution and a limited radiation length so that it can serve as an ideal inner tracker. Then a Cylindrical-GEM Inner Tracker (CGEM-IT) is proposed to be an option for the upgrade of the IDC for BES\uppercase\expandafter{\romannumeral3}. The proposed CGEM-IT consists of three independent layers of CGEM detectors with outer radii 87.5mm, 132.5mm and 175mm and lengths 532mm, 690mm and 847mm in axial direction. Each CGEM detector is composed by a cathode, three GEM foils and a two-dimensional readout anode in a concentric arrangement, as shown in Figure \ref{1.1}. High voltages are applied on each GEM foil full of micro bi-conical holes, providing strong electric fields in the holes to amplify the ionized electrons from a track to be measured. Such an inner tracker combined with ODC is expected to offer excellent spatial resolutions: about 100 $\mu$m in the r$\phi$ view and about 200 $\mu$m in the z direction, where r is the radius, $\phi$ is the azimuth angle and z is the coordinate along the central axis of a CGEM detector. The development of simulation and reconstruction softwares for CGEM-IT is essential to study its performance in track measurement and used for design optimization of the detector.

In this paper we will introduce the full simulation package for CGEM-IT, describe the implementation of the cluster reconstruction and track fitting algorithms for this detector in the BES\uppercase\expandafter{\romannumeral3} Offline Software System (BOSS) \cite{5} framework, and present the preliminary results on the resolutions of CGEM-IT and the performance of reconstructed tracks by using a Monte Carlo sample of single-muon events.

\section{Simulation of CGEM-IT}

Reliable Monte Carlo (MC) simulation plays an important role in describing the geometry and material of a detector and its response in track measurements. A full simulation package based on GEANT4 \cite{6} has been developed for CGEM-IT and the BOSS framework has been modified correspondingly.

\subsection{Simulation procedure}

Figure \ref{2.1} shows the flow chart of the simulation procedure. The geometries and materials of CGEM-IT including the micro holes in GEM foils are fully and exactly constructed according to the preliminary design of the CGEM-IT. Each GEM foil is a copper clad Kapton foil as thin as 54 $\mu$m with holes evenly distributed on it. The holes are of the same size with an external diameter of 70 $\mu$m. The space between the cathode and the first GEM foil is defined as the drift area, i.e. sensitive volume. The simulation is processed event by event. The final states (primary tracks) and their initial momenta of each event are produced by an event generator. 

To simulate the process that a charged primary track traverses the CGEM-IT, the track in drift area is propagated in discrete steps along the trajectories, during which all possible interactions between track and CGEM-IT are considered including ionization, Coulomb multiple scattering and so on. The information of each step in drift area is recorded as Cgem Hit. These Cgem hits are summed up to get the position and total energy loss of the track segment in the drift area as Cgem Truth. Then an algorithm called Cgem Digitizer converts the Cgem Truth to the electronic readout signal named as Cgem Digi. 

\begin{center}
  \includegraphics[width=0.5\textwidth]{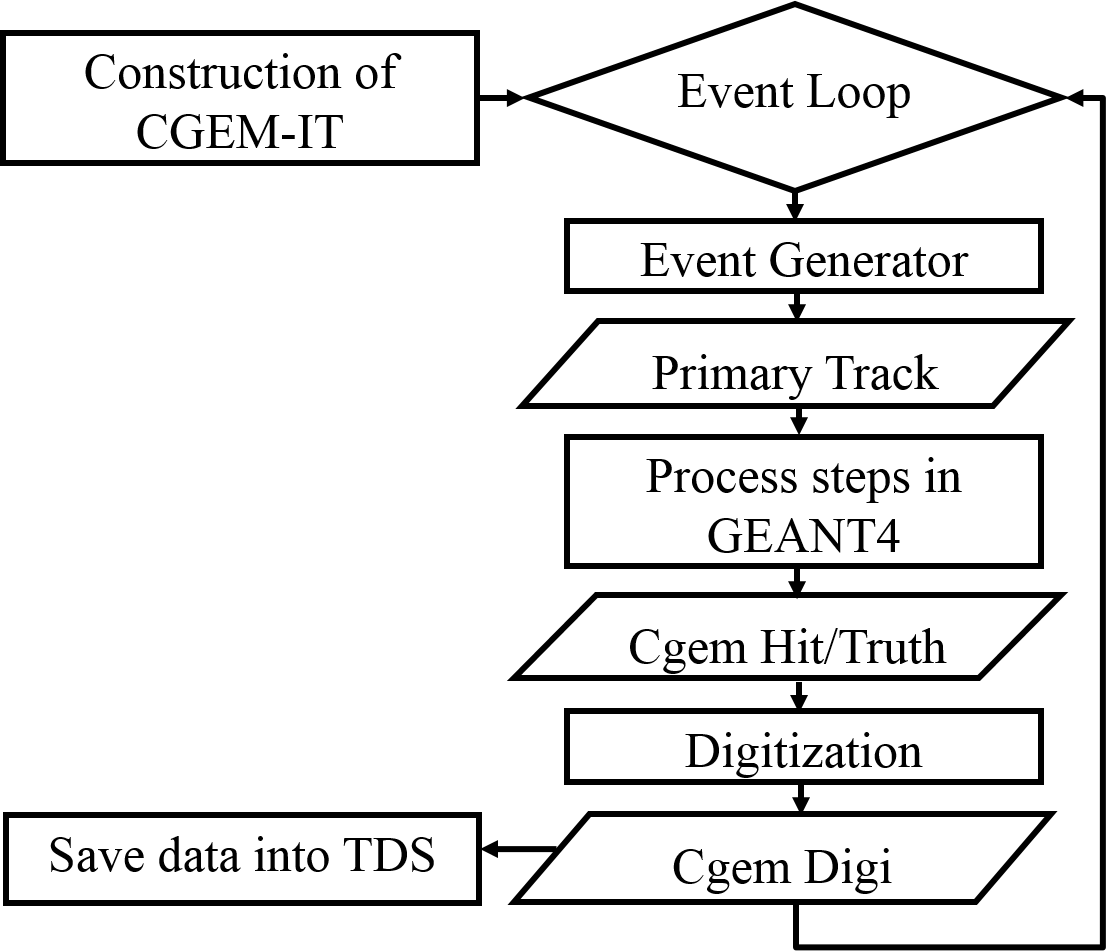}	
  \figcaption{\label{2.1} Flow diagram for the simulation of CGEM-IT}
\end{center}
\subsection{Digitization}

The electrons ionized by a charged particle crossing the drift region will be multiplied in the micro holes of the GEM foil and drift towards the anode where signals are induced. Three GEM foils are assembled to multiply the number of the electrons by a factor up to $10^6$\cite{3}. 

Both the magnetic field and diffusion of the electrons have impact on the drift properties and charge distribution in the readout anode \cite{7}. 

The presence of an axial magnetic field makes the drift line deviate from the direction of the electric field and the deviated angle is defined as Lorentz angle, as shown by the dashed lines in Figure \ref{1.1}. The effect of diffusion will enlarge the cluster of multiplied electrons. In this work, a simplified digitization model is used, in which the Lorentz angle $\alpha$ and electron diffusion are not considered. As shown by the solid line in Figure \ref{1.1}, the entrance and exit points in the drift area are projected to the anode plane as the extent of the signal in the readout anode. The readout anode of each layer is segmented with 650 $\mu$m pitch XV patterned strips with a stereo angle that changes depending on the layer geometry. The fired strips are determined by the size of the cluster of multiplied electrons. The charge deposited in the drift area is multiplied by an expected gain and shared by the fired strips. The identifier of each fired strip as well as its collected charge is recorded as one Cgem digi just like real experimental data. 

There are two different strip readout methods, the binary one and the analog one \cite{3}, lead to different digitization models. The binary readout method sets a fixed threshold on the total charge of the fired strips and the reconstructed position is the geometrical center of these strips. In the analog readout method,

the distributions of charge deposited along two readout coordinated will be recorded. The collected charge of each strip is obtained by the integration of the distributions so that one can set a threshold on the single strips and reconstruct the charge centroid of fired strips as the position of the cluster, which results in a better spatial resolution compared with the binary readout method. The precise shape of the charge distributions will be determined by the beam test on the CGEM planar prototype afterwards. So in this paper, the digitization model is based on the binary readout method, where total deposited charge are shared equally by the fired strips.

\subsection{Modification of framework}

As CGEM-IT is a new detector added to BES\uppercase\expandafter{\romannumeral3}, the BOSS framework doesn't provide necessary interfaces for the softwares related to CGEM-IT. Therefore we update the framework by adding the components for data processing with CGEM-IT. The information of Cgem Digi and Cgem Turth is stored in Transient Data Store (TDS) and can be accessed by the reconstruction algorithms via the event data service.

\section{Cluster reconstruction for CGEM-IT}

When the multiplied electrons reach the readout anode, a group of X and V strips will be fired. One cluster is defined in the anode plane and consists of continuous fired strips in X direction and V direction that have intersections. Figure \ref{3.1} exhibits one example of the composition of one cluster. Next we will discuss the cluster reconstruction and the obtained spatial resolution.

\begin{center}
	\includegraphics[width=0.5\textwidth]{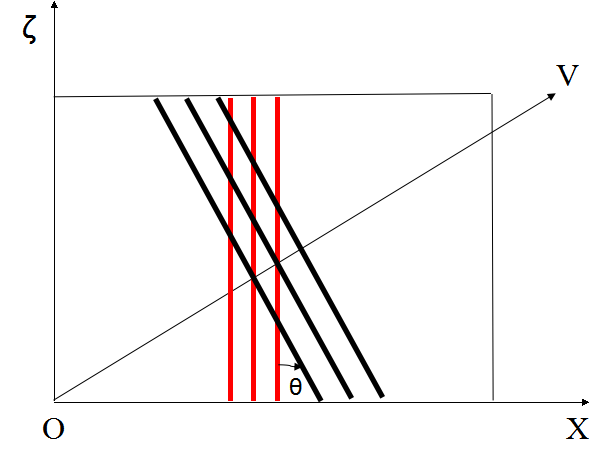}	
	\figcaption{\label{3.1} Schematic of one cluster induced by a charged track, composed of 3 X-strips and 3 V-strips on an unrolled anode plane, where the stereo angle $\theta$ is the angle between X-strip and V-strip, the X-axis is  perpendicular to X-strips, the V-axis is perpendicular to V-strips and the $\zeta$-axis is parallel to X-strips.}
\end{center}

\subsection{Cluster reconstruction algorithm}

Firstly, the information of fired strips is retrieved from the raw data, including the strip ID, strip type (X or V) and collected charge. Then sets of continuous fired strips in X and V directions are found and defined as X-clusters and V-clusters respectively. Next the possible intersections between X-clusters and V-clusters are searched for and saved as XV-clusters. The position in X (V) of each cluster is acquired by calculating the charge weighted geometrical center of the fired strips that compose this cluster:
\begin{eqnarray}
	X & = & \frac{\sum_{i=1}^{N_X}{X_i{Q_i}}}{\sum_{i=1}^{N_X}{Q_i}}	    
\end{eqnarray}
\begin{eqnarray}
	V & = & \frac{\sum_{i=1}^{N_V}{V_i{Q_i}}}{\sum_{i=1}^{N_V}{Q_i}}	  
\end{eqnarray}
where $i$ is the index of strips, $X_i$ $(V_i)$ is the middle position of strip $i$, $Q_i$ is the charge collected by each strip and $N_X$ $(N_V)$ is the number of fired X-strips (V-strips). 
The corresponding position in the absolute cylindrical coordinates $(r,\phi,z)$ is given by 
\begin{equation}
\left\{
\begin{array}{r@{\;=\;}l}
r & R\\
z & (V-X{\cos\theta})/sin\theta \\
\phi & X/R
\end{array}
\right.
\end{equation}
where R is the radius of the anode in which the cluster is found, $\theta$ is the stereo angle between X-strips and V-strips. To obtain the middle positions where charged particles pass the CGEM in the drift area, a correction to the reconstructed positions of clusters is demanded. According to the simplified digitization model, the position of a track in drift area has the same azimuth angle $\phi$ and the same $z$ as the cluster, but with a different radius. The corresponding output data model is defined by a class RecCgemCluster, including the following information: layer ID, GEM ID, cluster type (X, V or XV), collected charge and reconstructed position.

\subsection{Spatial resolution for CGEM-IT clusters}

The spatial resolution for CGEM-IT clusters is an essential performance indicator of the cluster reconstruction algorithm. The residuals in X and V directions are defined as:
\begin{align}
\delta_X & =X_{rec}-X_{truth} \\
\delta_V & =V_{rec}-V_{truth}
\end{align}
where $X_{rec}$ and $V_{rec}$ are the reconstructed positions in the drift layer in X-direction and V-direction while $X_{truth}$ and $V_{truth}$ are the true positions. For a Monte Carlo sample of $\mu^-$ with a momentum of 1.0 GeV/c, the distribution of the residuals in X and V are shown in Figure \ref{3.2}. Note that the distribution of residuals in X is not a pure Gaussian, but rather a superposition of rectangular and Gaussian. The Gaussian part (filled region in Fig.\ref{3.2}(a))comes from clusters with multiple strips while the rectangular part (dashed line in Fig.\ref{3.2}(a)) comes from clusters with single strip. The RMS (Root Mean Square) of this curve is 217.1 $\mu$m and defined as the resolution in X-direction. A fit of the distribution of residuals in V to a Gaussian function gives a resolution of 129 $\mu$m.

\begin{center}
	\includegraphics[width=0.5\textwidth]{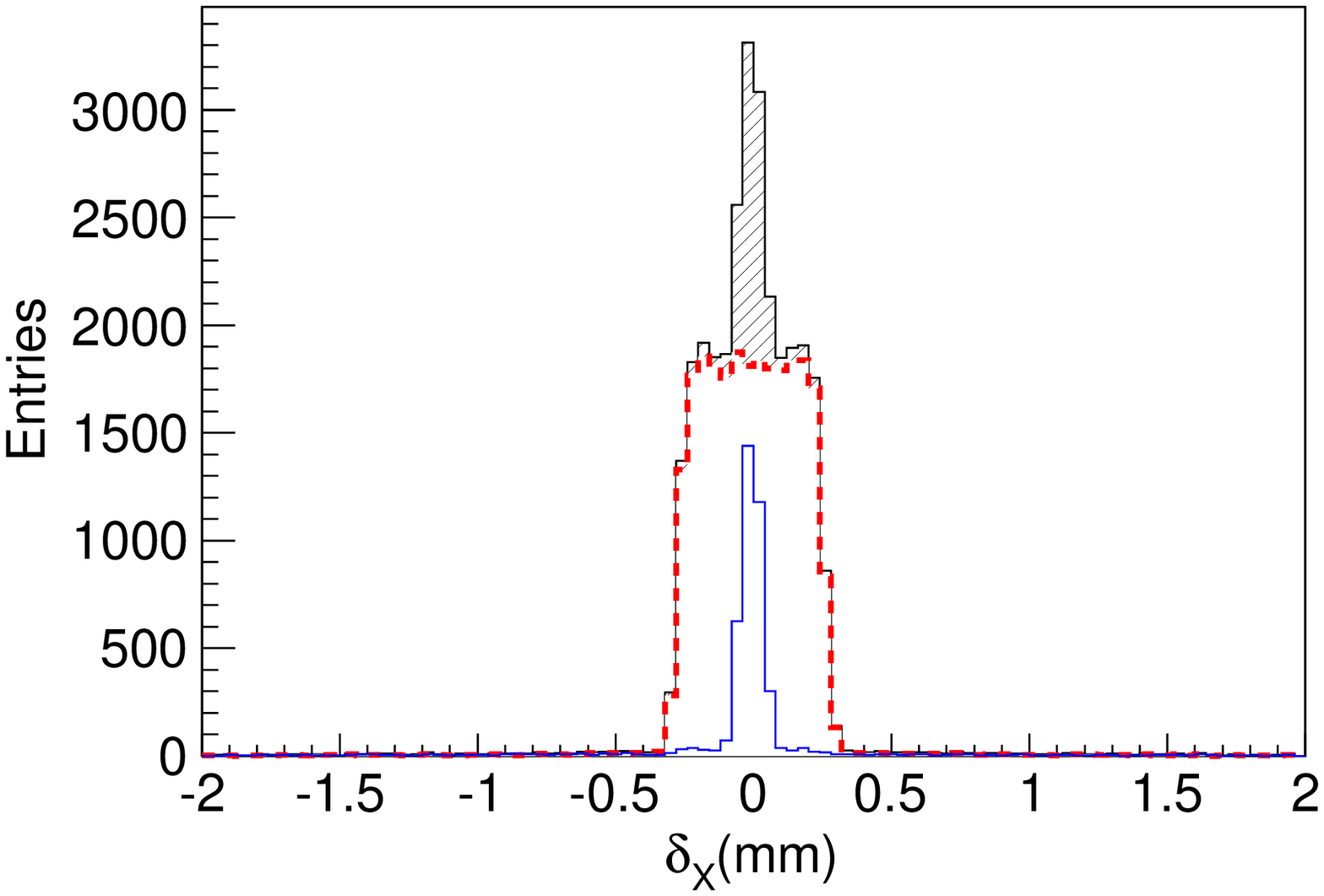}\put(-165,130){(a)}
	
	\includegraphics[width=0.5\textwidth]{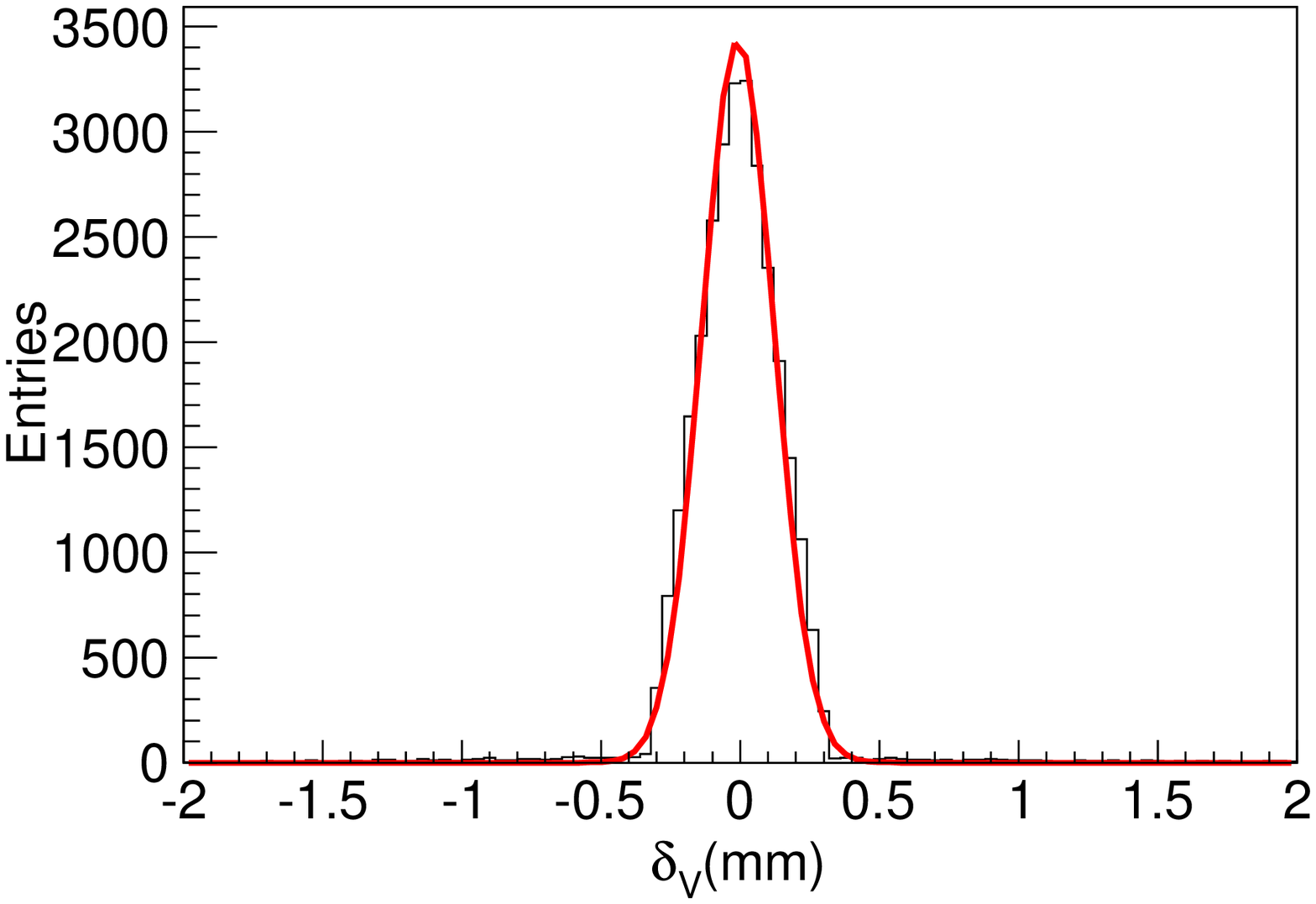}\put(-165,130){(b)}	
	\figcaption{\label{3.2} The position residuals in X (a) and V (b) of the cluster for a Monte Carlo sample of single 1.0 GeV/c $\mu^-$ tracks. The contribution from clusters with multiple strips (filled region) and the part from clusters with single strip (dashed line) are shown in (a). The solid line in (b) is the Gaussian fit curve for the residuals distribution in V-direction.}
\end{center}

\section{Track fitting by Kalman filter with CGEM-IT clusters}

In a uniform magnetic field a charged particle moves along a helical trajectory. 
So a track can be described by a helix model with 5 parameters $(d_\rho, \phi_0, \kappa, d_z, \tan\lambda)$ \cite{8}, where $d_\rho$ and $d_z$ are the signed distance of closest point to the pivot in $x-y$ plane and z direction respectively, $\phi_0$ is the azimuth angle of the pivot to the helix center, $\kappa$ is $1/p_t$ with the same sign as the charge of the track and $\tan\lambda$ is the slope of the track.

The track finding algorithm \cite{8,9} and track fitting algorithm by Kalman filter \cite{10} in service using MDC hits can be easily adopted to accomplish the same task using ODC hits only. To combine the cluster measurements by CGEM-IT, the adopted track fitting by Kalman filter is extended to be able to extrapolate the track found with ODC hits into CGEM-IT and fit the track with these additional clusters in a right sequence.

In this section, we will discuss the extension of Kalman filter based track fitting to use CGEM-IT clusters and present the obtained resolutions for the track parameters after the track fitting.

\subsection{Implementation of the track fitting by Kalman filter for CGEM-IT}

The description of geometries and materials of CGEM-IT is implemented in the initialization of the track fitting algorithm and identical with that from simulation package.
 
The reconstructed clusters are retrieved as input. 

The track parameters and their covariance matrix in the ODC are propagated into CGEM-IT by many small steps, during which the effects from inhomogeneous magnetic field, energy loss and multiple scattering in the material are included. When the track is extrapolated into a drift area of CGEM-IT, the measured position from the CGEM-IT cluster and the predicted track parameters are combined to update the track parameters. This process is repeated for the subsequent regions, with each iteration yielding more accurate track parameters and their covariance matrix, and stops at the original point. The obtained track parameters are taken as the reconstruction results for this track.

\subsection{Vertex resolution}

To obtain the vertex resolution of the reconstructed tracks, the residuals of $d_\rho$ and $d_z$ are calculated which are the differences between the fited values and the truth at the generation level. Figure \ref{4.1} shows the distribution of the residuals for the Monte Carlo sample single $\mu^-$ with 1.0 GeV/c, and a Gaussian fit yields a resolution of 192 $\mu$m in $d_\rho$ and 0.33mm in $d_z$.
\begin{center}
	\includegraphics[width=0.5\textwidth]{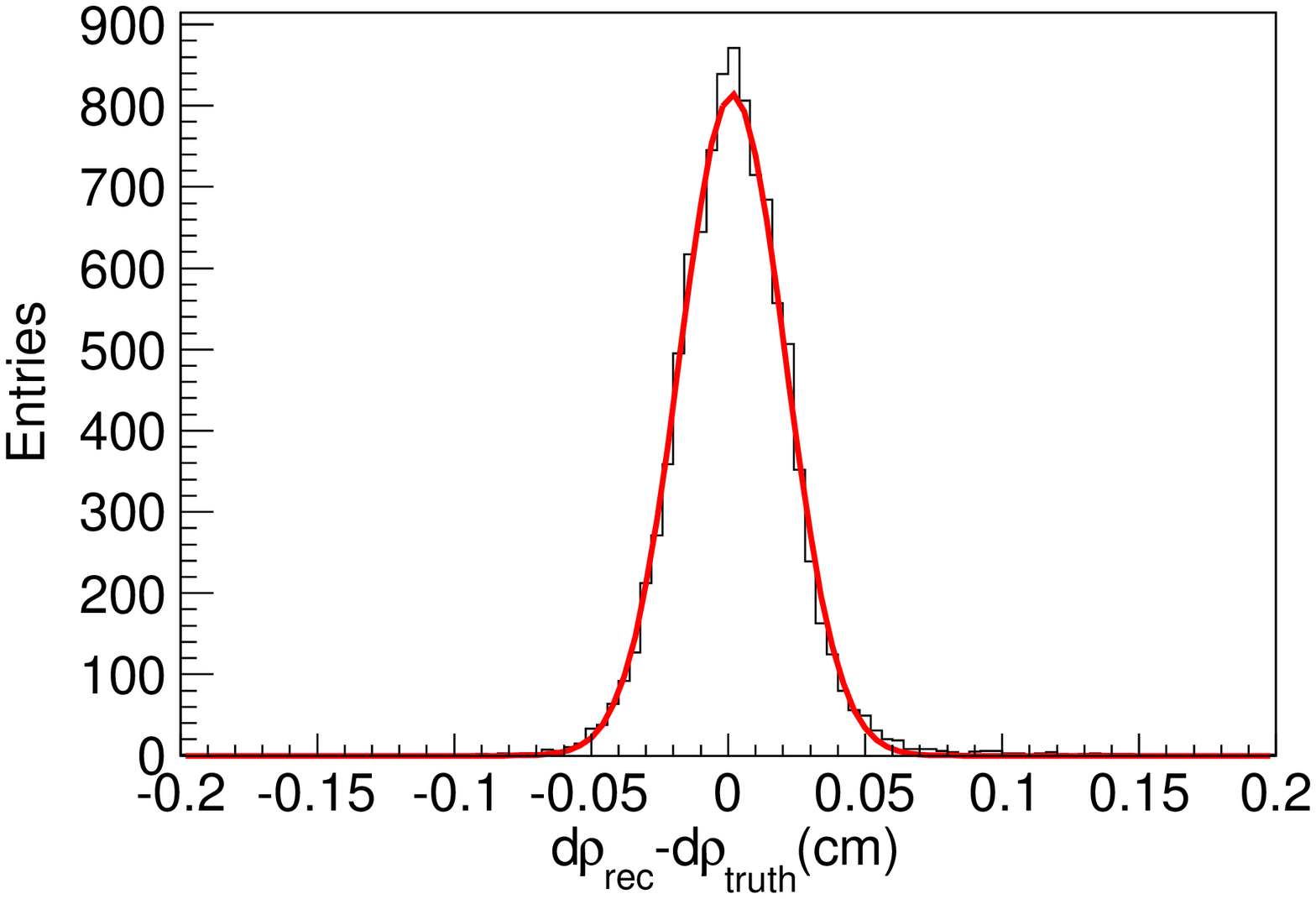}\put(-165,130){(a)}
	
	\includegraphics[width=0.5\textwidth]{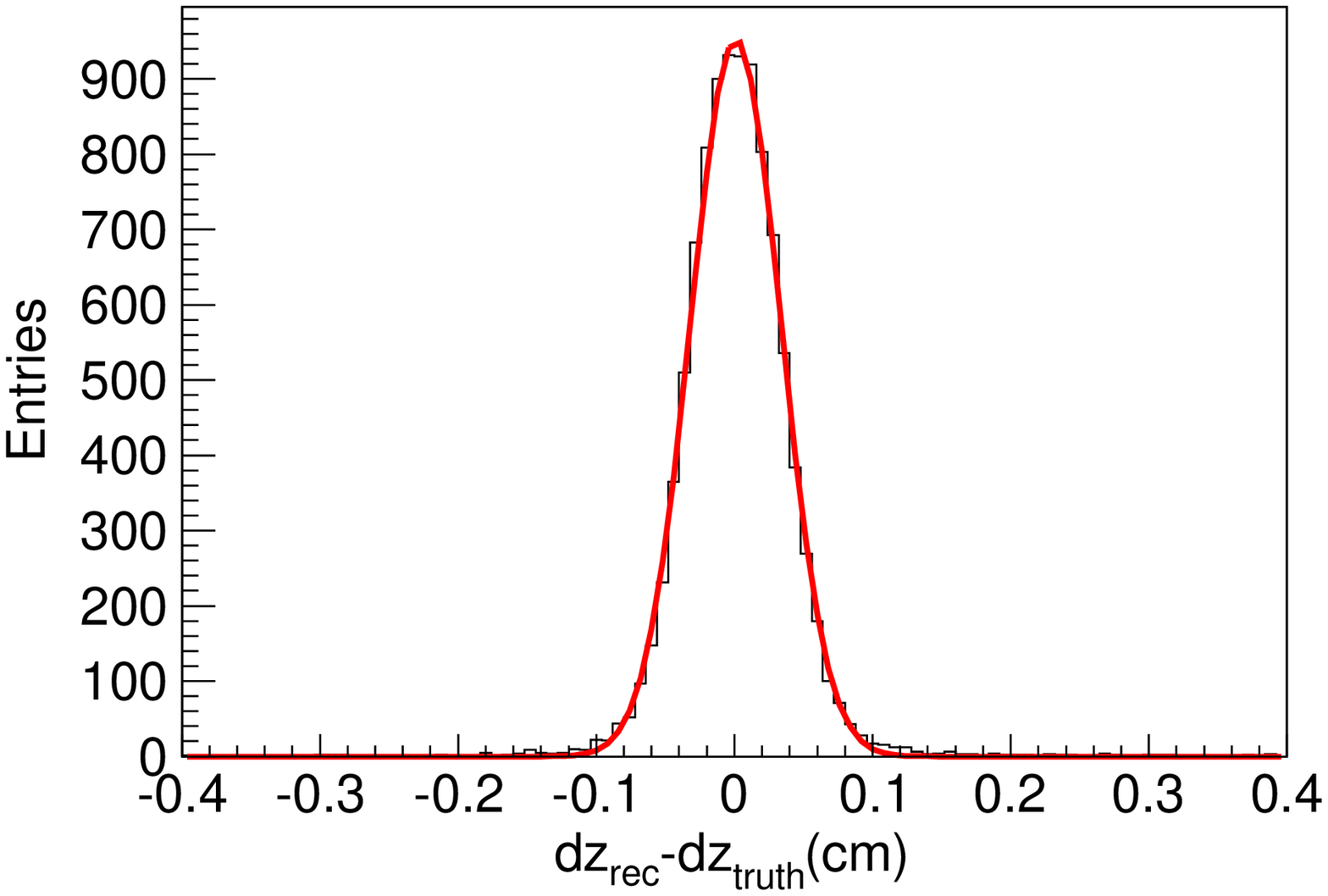}\put(-165,130){(b)}
	\figcaption{\label{4.1} The $d_\rho$ (a) and $d_z$ (b) residuals for single 1.0 GeV/c $\mu^-$ tracks with CGEM-IT. The curves are the results of a Gaussian fit.}
\end{center}
The vertex resolution in $d_\rho$ and $d_z$ for single $\mu^-$ tracks with different momenta for CGEM-IT and IDC is displayed in Figure \ref{4.2}. These results reflect that a tracking system using the CGEM-IT has a comparable $d_\rho$ resolution and a significantly improved $d_z$ resolution relative to the case using IDC.

\begin{center}
	\includegraphics[width=0.5\textwidth]{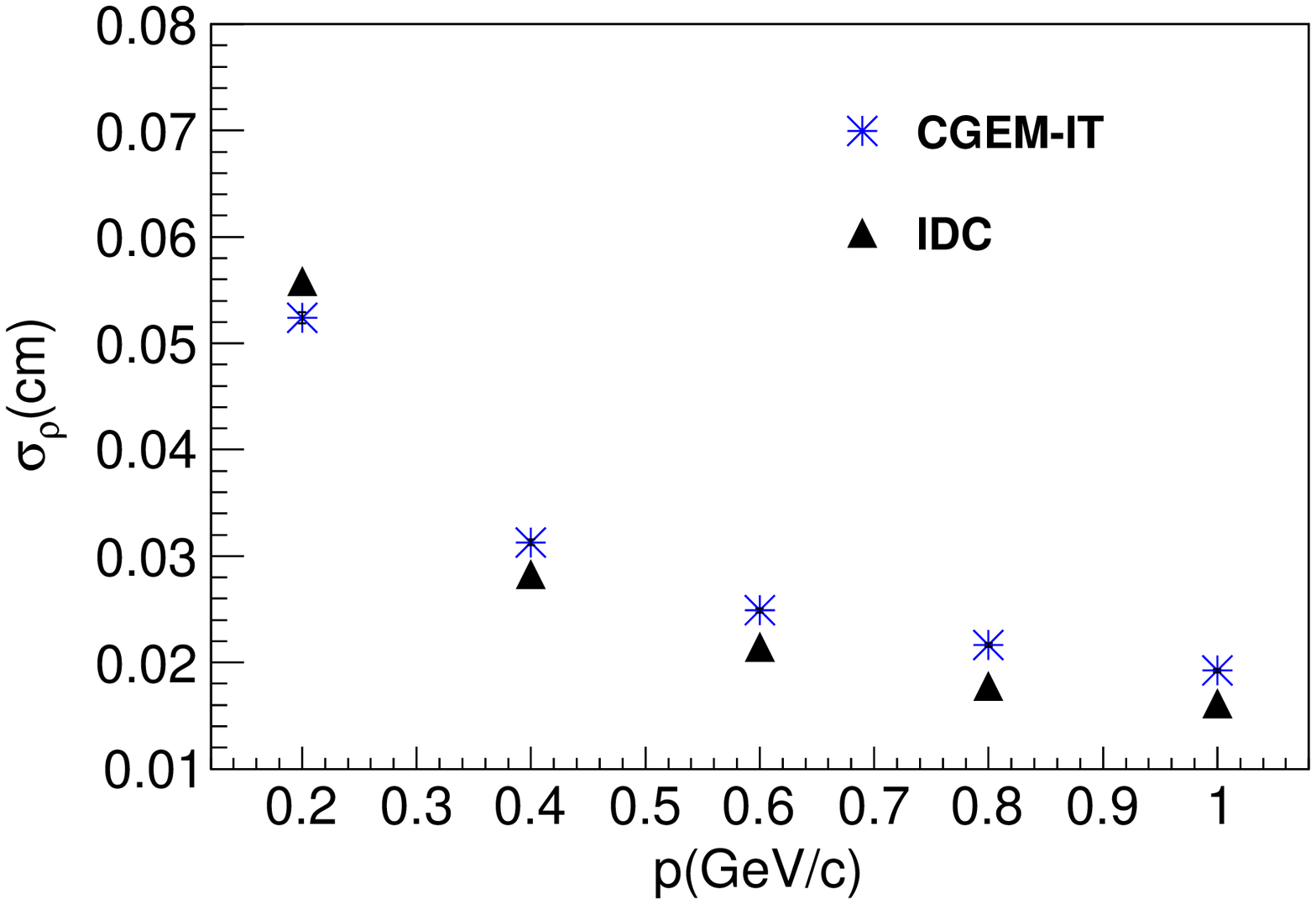}\put(-165,130){(a)}
	
	\includegraphics[width=0.5\textwidth]{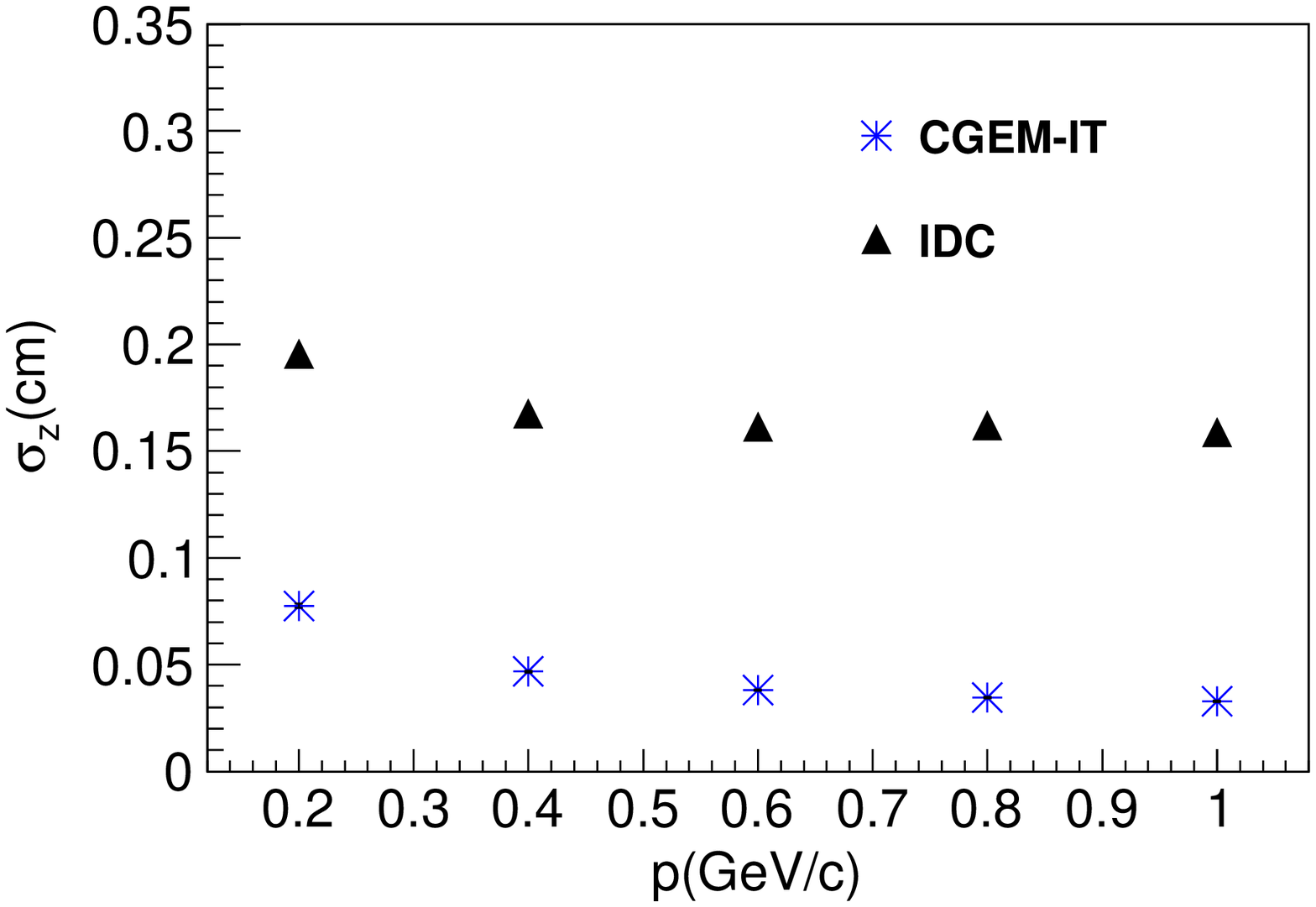}\put(-165,130){(b)}
	\figcaption{\label{4.2} The $d_\rho$ (a) and $d_z$ (b) resolutions of single $\mu^-$ tracks with different momenta for CGEM-IT and IDC.}
\end{center}
\subsection{Momentum resolution}

The momentum resolution is represented by the distribution of the momentum residual, which is defined by:
\begin{equation}
\delta_p=p_{rec}-p_{truth}
\end{equation}
where $p_{rec}$ is the reconstructed momentum at interaction point and $_{truth}$ is the momentum at the generation level. The $\delta_p$ distribution curve for 1.0 GeV/c muons is shown in Figure \ref{4.3} as an instance. This curve is fitted by a Gaussian function and the obtained standard deviation 5.5 MeV/c is regarded as the resolution of momentum.
\begin{center}
	\includegraphics[width=0.5\textwidth]{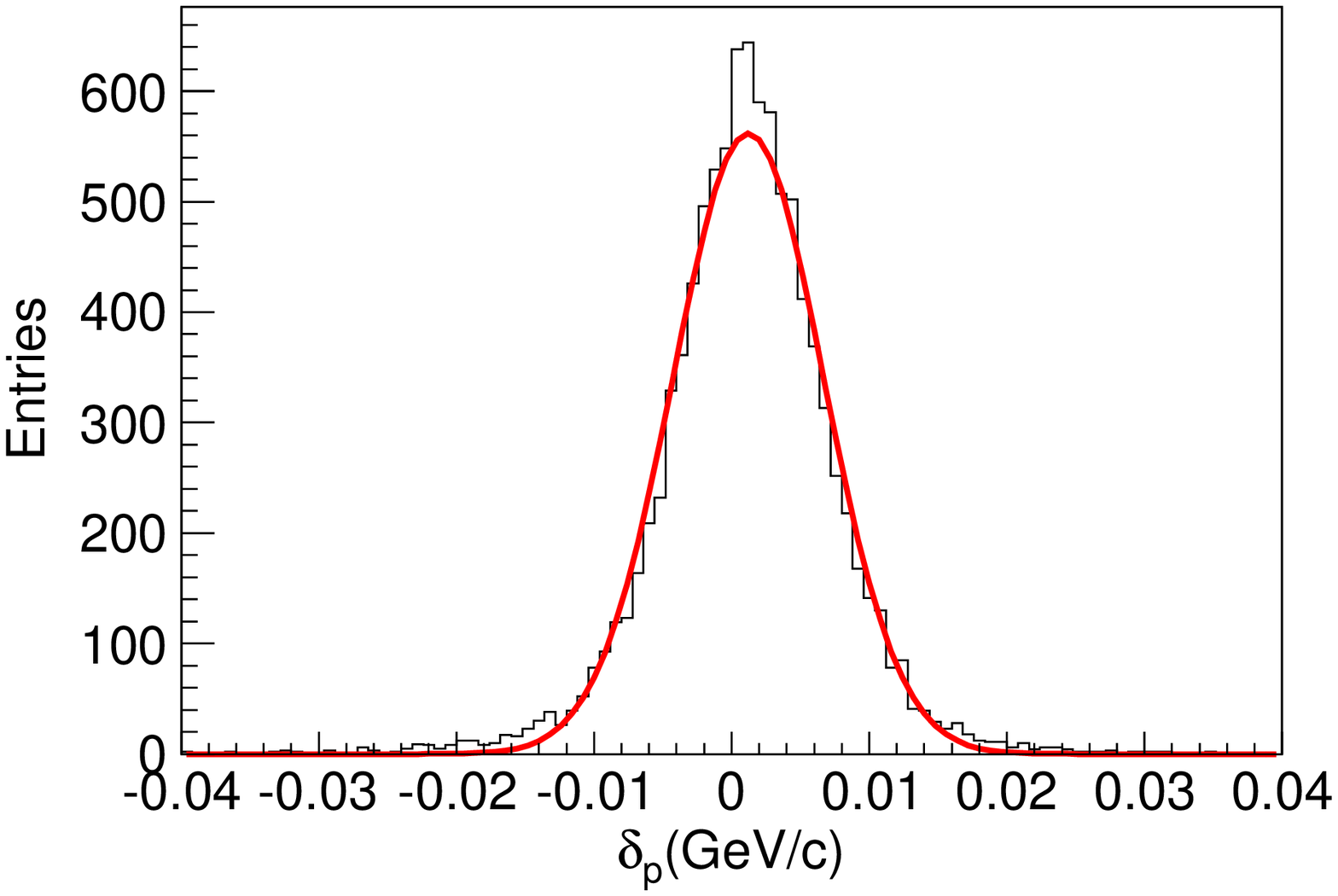}	
	\figcaption{\label{4.3} The distribution of momentum residual for 1.0 GeV/c $\mu^-$ tracks is fitted by a Gaussian function (red solid line).}
\end{center}
Figure \ref{4.4} summarizes momentum resolutions for single muon samples simulated with different momenta for CGEM-IT and IDC. These figures illustrate that their momentum resolutions are at a comparable level. CGEM-IT has a larger material budget and offers only three layers of measurements compared with IDC. However, the good spatial resolution in two dimensions of CGEM-IT compensates the disadvantages so that a comparable momentum resolution can be achieved. 
\begin{center}
	\includegraphics[width=0.5\textwidth]{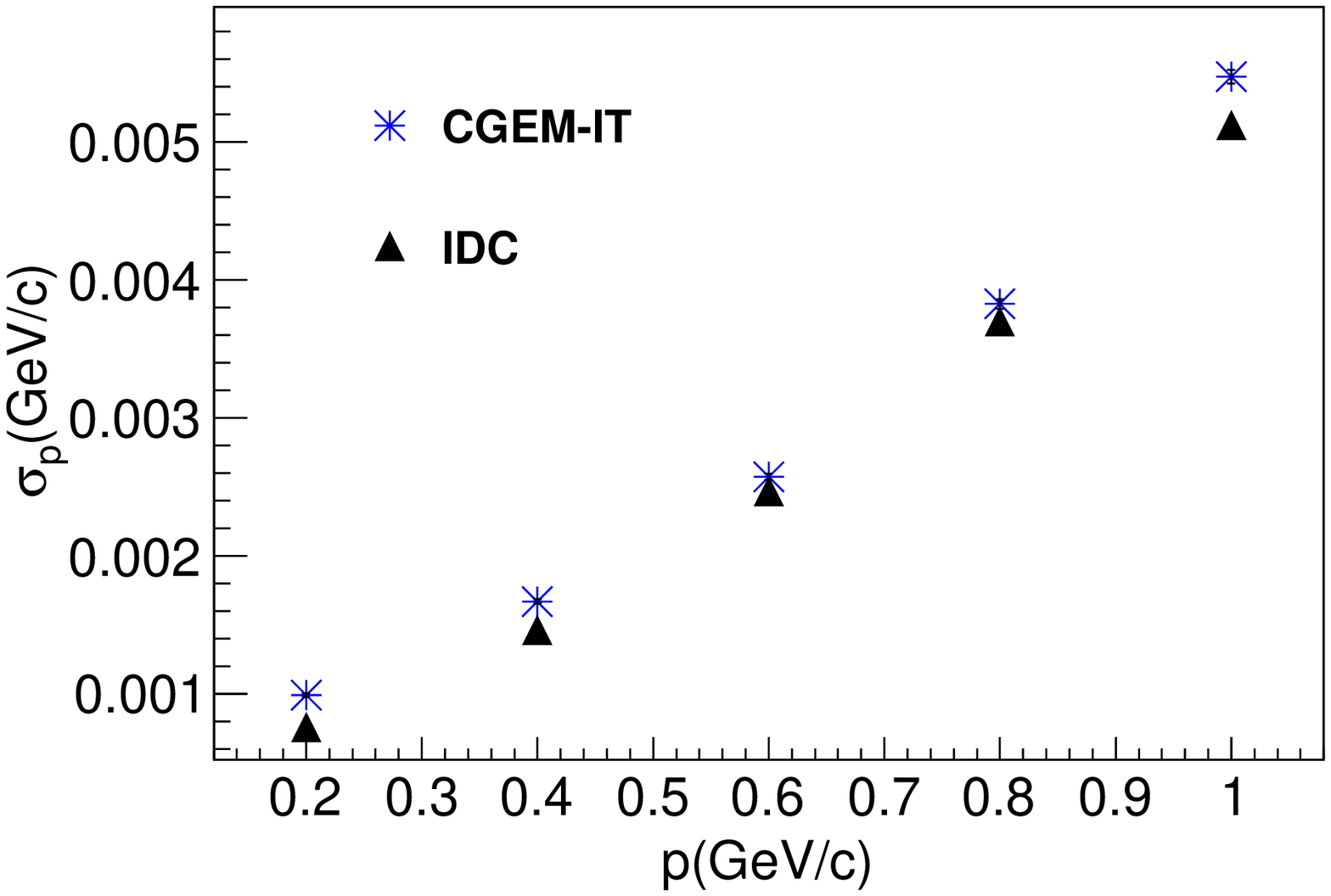}	
	\figcaption{\label{4.4} The momentum resolutions of the reconstructed muon at different momenta using full simulation of single $\mu^-$ events in CGEM-IT and IDC.}
\end{center}
\section{Summary and outlook}

In order to solve the aging problems of IDC in BES\uppercase\expandafter{\romannumeral3}, a CGEM-IT is suggested to be used to upgrade the IDC. In this paper, we present the implementation of the softwares for CGEM-IT. The general flow for CGEM-IT including simulation, cluster reconstruction and track fitting is realized. The results for a set of simulated $\mu^-$ events indicate that the vertex resolution in z direction is significantly improved by CGEM-IT and meanwhile the momentum resolution and vertex resolution in r direction stay at the same level with respect to IDC. The preliminary results demonstrate that the CGEM-IT and the implemented track reconstruction softwares can basically fulfill the track reconstruction task as a replacement of IDC. 

The implementation of a more detailed digitization model based on the analog readout method will improve the spatial resolutions of CGEM-IT. Further studies should also be performed to reach a better understanding of the detector, including developing the track segment finding algorithms for CGEM-IT and joint tracking algorithms for CGEM-IT and ODC, optimizing the reconstruction softwares under the environment with noise and background.

\end{multicols}

\vspace{-1mm}
\centerline{\rule{80mm}{0.1pt}}
\vspace{2mm}

\begin{multicols}{2}

\end{multicols}

\end{CJK*}
\end{document}